\documentstyle[epsfig,12pt]{article} \textwidth=16cm
\begin{document}
\title{The electromagnetic instabilities propagation in weak relativistic quantum plasmas}
\author{ M. Mahdavi\thanks{Email:{\tt m.mahdavi@umz.ac.ir}}
$~$and H. Khanzadeh\thanks{Email:{\tt h.khanzadeh@stu.umz.ac.ir}}}
\date{{\small Physics Department, University of Mazandaran , P. O. Box
47415-416, Babolsar, Iran}\\} \maketitle
\begin{abstract}
The electromagnetic instabilities excited by the temperature
anisotropy have been always one of the interesting issues in real
high-density physical systems, where the relativistic and quantum
effects due to spin can be important. This paper discusses the case
where plasma is not strongly coupled but is still in regimes where a
classic plasma description is not fully adequate. The length scale
of the plasma can be larger than the de-Broglie length so that the
quantum effects relevance to the spin can be significant. In
addition, the only relativistic effects are due to the electrons
spin, for example the spin-orbit coupling effect (the weak (semi)
relativistic effects). Obtained results imply that these effects can
not be important in the ICF subjects while these can lead to
significant results in the astrophysical subjects because of the
strong magnetic fields. It is found that the weakly relativistic
effects can cover the magnetic dipole force and the spin precession
effects so that the growth rate of the instability can increase
compared to the non relativistic spin polarized cases. Indeed, it is
expected that the probability of electron capture in the background
magnetic fields and the particle energy dissipation will be reduced
so that there will be
a high portion of free energy in the system.\\
{\textbf{ Key words:} Spin, Relativistic effects, Spin-Orbit
coupling, Electromagnetic instability, Temperature anisotropy.}
\end{abstract}
\newpage

\section{Introduction}
Plasma includes a highly complex physical system where there are a
wide variety of theoretical methods with a common general principle
to introduce them. It can be expected that one method can be
transferred between different plasma system, for example transfer
method for treating nonlinearities in classical plasmas to quantum
plasmas [1]. In this order, lowest order corrections have been
applied in terms of nonlocal terms related to the tunneling aspects
of the electrons. In other words, the electron spin (the possibility
of large-scale magnetization of plasma) can be effective on the
dynamics of classical systems where the spin waves can be excited by
flux of the intensity neutrons in dens magnetized plasmas [2]. In
addition, the relativistic effects may have significant effects,
too. In this case, for example in laboratory plasmas such as laser
generated plasmas and in nature in particular for planetary
interiors and stars [3], combining the quantum relativistic effects
need more complex dynamic methods for obtaining accurate
descriptions of a host of phenomena. There have been more studies
related to the quantum effects due to the particle dispersive and
the Fermi pressure where the magnetization current and dipole force
due to the electron spin are included [2,4-6]. Studies show that,
the spin quantum hydrodynamic theories (for example fluid models
like MHD or two-fluid models) are not relevant to conventional
laboratory or astrophysical plasmas while the kinetic theory needs
more exact studies [7]. One of the most accurate theories is based
on the Kandanoff-Bay kinetic equations [8]. This theory (containing
memory effects (non local terms) both in space and time) can be
effective even on time scale shorter than the typical relaxation
time of system but is not specially adopted to some of the
applications for example, a high-intensity laser-plasma interaction
and high-energy density physics. Therefore, here, we are interested
in a method similar to Asenjo et al. [9] where plasmas are not
strongly coupled but still in regimes where, a classical plasma
description is not fully adequate. Notice that, increasing effects
of the magnetic fields can add more phenomena to plasma-wave
problems where anisotropy problem is one of the most important.
Therefore, here, we are particularly interested in velocity space
instabilities excited by the temperature anisotropy in weak
relativistic spin-1/2 quantum plasmas. In this order, in section 2,
first the dispersion relation and as a result, the growth rate of
the instabilities are derived in presence of the weak (semi-)
relativistic effects. A comparison of results to our previous study
will be made in section 3 for two real physical systems ie. ICF and
astrophysical plasmas and finally, our main conclusions are
summarized in section 4.
\section{Basic theory}
Introducing suitable evolution equation is based on the Asenjo et
al. work [9] where a semi-relativistic transformation is introduced
by applying the Foldy-Wouthuysen transformation for particle in
external fields [9,10]:
\begin{eqnarray}
  &\hat{H}=mc^{2}+q\phi+\frac{1}{2m}\left(\hat{p}-\frac{q}{c}A\right)^{2}-\frac{q\hbar}{2mc}\sigma\cdot B& \nonumber\\
  &+\frac{\hbar^{2}q}{8m^{2}c^{2}}\nabla\cdot E-\frac{\hbar q}{4m^{2}c^{2}}\sigma\cdot\left[E\times\left(\hat{p}-\frac{q}{c}A\right)\right]& \nonumber\\
  &-\frac{i\hbar^{2}q}{8m^{2}c^{2}}\sigma\cdot\nabla\times E+\frac{1}{8m^{3}c^{2}}\left(\hat{p}-\frac{q}{c}A\right)^{4}&
\end{eqnarray}
where $m$ is the electron mass, $q=-e$ is the electron charge, $c$
is the speed of light and $\hat{p}$, $\phi$ and $A$ are the momentum
operator, the scalar and vector potential, respectively. The
quantity $\hbar$ is the reduced Planck constant, $\sigma$ denotes a
vector containing the $2\times2$ Pauli matrices and $B$ and $E$ are
the magnetic and electric field, respectively. The first four terms
constitute the Pauli Hamiltonian while the fifth and eighth terms
are the Darwin and mass-velocity correction terms, respectively. The
sixth and seventh terms together give Thomas precession and
spin-orbit coupling due to higher order corrections, too.

It is well known, plasma is introduced in a quantum regime where the
most suitable kinetic model is the Wigner model [11] (in the absence
of interaction between particles) when the characteristic de-Broglie
wave length is equal to the Fermi wave length. The quantum effects
associated with the particle dispersive effects disappear at the
spatial length-scale higher than the de-Broglie length while the
only reminded quantum effects are relevant to the spin of particle.
Here, according to this, the evolution equation can be found for
above Hamiltonian using the Wigner transformation in phase space and
the Q-transformation in spin space as [9]:
\begin{eqnarray}
  &\frac{\partial f}{\partial t}+\left\{\frac{p}{m}+\frac{\mu}{2mc}E\times\left(s+\nabla_{s}\right)\right\}\cdot\nabla_{x}f& \nonumber\\
  &+q\left(E+\frac{1}{c}\left\{\frac{p}{m}+\frac{\mu}{2mc}E\times\left(s+\nabla_{s}\right)\right\}\times B\right)\cdot\nabla_{p}f& \nonumber\\
  &+\frac{2\mu}{\hbar}s\times\left(B-\frac{P\times E}{2mc}\right)\cdot\nabla_{s}f+\mu\left(s+\nabla_{s}\right)  \cdot\partial_{x}^{i}\left(B-\frac{P\times E}{2mc}\right)\partial_{p}^{i}f& \nonumber\\
  &-\frac{\hbar^{2}q}{8m^{2}c^{2}}\partial_{x}^{i}\left(\nabla\cdot E\right)\partial_{p}^{i}f=0&
\end{eqnarray}
here, the only terms up to first order in the velocity is kept and
the gamma factor is put to unity where the relation between the rest
frame spin, $s$, and the spatial part of the spin four-vector, $S$,
implies on
$S=s+\left[\gamma^{2}/\left(\gamma+1\right)\right]\left(v\cdot
s\right)v/c^{2}$. The quantity $\mu=\frac{\hbar qg}{4mc}$ is the
intrinsic magnetic moment with the spin factor $g=2.00232$. Notice
that, the Hamiltonian (1) and the Dirac theory have been started
from the exactly value $g=2$ while we will use the value $g=2.00232$
in next. This may suggest that the corrections (the QED corrected
value [12]) should be added to Hamiltonian. It can add new terms to
the evolution equation so that, extra terms are smaller than those
kept by a factor of the order $\left(g-2\right)$. Since, the
QED-corrections to the Dirac Hamiltonian will not be include in
theory otherwise modifying the spin $g$-factor value. The
distribution function, $f$, is sum of the time-independent
unperturbed and time-dependent perturbed distribution functions as
$f = f_{0} + f_{1}$, where it is normalized to the number density.
In the absence of the background electric field and in the presence
of the background magnetic field, such as $B = B_{0} + B_{1}$, Eq.
(2) can be rewritten as:
\begin{eqnarray}
&\frac{\partial f_{1}}{\partial t}+\frac{p}{m}\cdot\nabla_{x}f_{1}+\frac{q}{mc}p\times B_{0}\cdot\nabla_{p}f_{1}
+\frac{2\mu}{\hbar}s\times B_{0}\cdot\nabla_{s}f_{1}=&\nonumber\\
&-q E_{1}\cdot\nabla_{p}f_{0}-\frac{q}{mc}p\times B_{1}\cdot\nabla_{p}f_{0}-\mu\nabla_{xi}\left[B_{1}\cdot\left(s+\nabla_{s}\right)\right]\nabla_{pi}f_{0}&\nonumber\\
&+\frac{\mu}{2mc}\nabla_{xi}\left[\left(p\times E_{1}\right)\cdot\left(s+\nabla_{s}\right)\right]\nabla_{pi}f_{0}
-\frac{q\mu}{2mc^{2}}\left[E_{1}\times\left(s+\nabla_{s}\right)\right]\times B_{0}\cdot\nabla_{p}f_{0}&\nonumber\\
&-\frac{2\mu}{\hbar}s\times B_{1}\cdot\nabla_{s}f_{0}+\frac{\mu}{\hbar mc}s\times\left(p\times E_{1}\right)\cdot\nabla_{s}f_{0}&
\end{eqnarray}
where the last term in Eq. (2) is ignored because of the smaller
contribution compare to other terms. In fact, we can ignore the two
and higher orders of the Plank constant because of the applied
assumption (the scale length larger than the de-Broglie length).

Notice that, different quantum effects can be contained in the
unperturbed distribution function, such as Fermi-Dirac statistics,
Landau quantization, and spin-splitting of the energy states [6,
13]. The electrons’ behavior as a degenerate electron gas follows
the known Fermi-Dirac statistics in plasmas with low temperature and
high density. The Landau quantization (quantization of the
perpendicular energy states) is important in the regime of strong
magnetic fields or very low temperatures (when
$\frac{\hbar\omega_{c}}{K_{B}T}\rightarrow 1$ with the electron
cyclotron frequency, $\omega_{c}$, and the Boltzmann constant,
$K_{B}$). In the presence of spin quantum effects, the unperturbed
distribution function must contain the properties of the spin space.
In this situation, there are different spin states with different
probability distributions for orienting the particles magnetic
moment in the background magnetic field. For cases where the
chemical potential, $\mu_{c}$, is large and the difference between
the nearby Landau levels is smaller than the thermal energy, the
velocity (momentum) distribution approaches the classic Maxwellian
distribution, while the remaining quantum effect is due to the
probability distribution of the spin-up and spin-down population
states. Thus, the distribution can be approximated by [14]:
\begin{equation}
  f_{0} =\sum_{\nu=\pm1}F_{0\nu}\left(p\right)\left(1 + \nu cos\theta_{s}\right)
\end{equation}
where the function $F_{0\nu}$ is the normalized function of the
momentum variables and the indexes $+$ and $-$ define the spin-up
and the spin-down states, respectively. Now, let us assume that the
background magnetic field is static, homogeneous, and points in the
$\hat{z}$-direction, ie. $B_{0} = B_{0}\hat{z}̂$ and the unstable
electromagnetic waves propagate in the direction of the z-axis, such
that the wave vector will be defined as $\vec{K}=k_{z}\hat{z}$.
Therefore $f_{1}$ can be expanded in eigenfunctions to the operator
of the right-hand side as:
\begin{equation}
  \tilde{f}_{1}=\frac{1}{2\pi}\sum_{a,b=-\infty}^{\infty}g_{a,b}(p_{\bot}, p_{z}, \theta_{s})e^{-i\left(a\phi_{p}+b\phi_{s}\right)}+c.c.
\end{equation}
where $c.c.$ stands for complex conjugate and a standard ansatz of
quasi-monochromatic harmonic variation on the perturbed quantities
is used, ie.
$\textbf{E}_{1}=\tilde{\textbf{E}}_{1}e^{i\left(\vec{K}\cdot\vec{X}-\omega
t\right)}$,
$\textbf{B}_{1}=\tilde{\textbf{B}}_{1}e^{i\left(\vec{K}\cdot\vec{X}-\omega
t\right)}$ and
$f_{1}=\tilde{f}_{1}e^{i\left(\vec{K}\cdot\vec{X}-\omega t\right)}$.
Linearizing Eq. (3) can produce an exact form of the function
$g_{ab}$ where it will result to:
\begin{eqnarray}
&\tilde{f}_{1}=\sum_{b=\pm1}\frac{\mu i}{\hbar}\left(\tilde{B}_{1y}-\frac{p_{z}\tilde{E}_{1x}}{2mc}
-\frac{ib}{2mc}p_{z}\tilde{E}_{1y}cos2\theta_{s}-ib\tilde{B}_{1x}\right)&\nonumber \\
&\times\frac{\partial f_{0}}{\partial\theta_{s}}
\frac{e^{-ib\phi_{s}}}{\left(\omega-k_{z}p_{z}/m-b\omega_{cg}\right)}&\nonumber \\
&-\sum_{b=\pm1}\frac{\mu i}{4mc}\left(bp_{z}\tilde{E}_{1x}+ip_{z}\tilde{E}_{1y}-2mcb\tilde{B}_{1y}+i2mc\tilde{B}_{1x}\right)
k_{z}cos\theta_{s}&\nonumber \\
&\times\frac{\partial^{2}f_{0}}{\partial\theta_{s}\partial p_{z}}\frac{e^{-ib\phi_{s}}}{\left(\omega-k_{z}p_{z}/m-b\omega_{cg}\right)}&\nonumber \\
&+\sum_{b=\pm1}\frac{\mu i}{4mc}(bp_{\bot}\tilde{E}_{1x}cos\left(\theta_{s}\right)-bp_{z}\tilde{E}_{1x}sin\left(\theta_{s}\right)
-ip_{z}\tilde{E}_{1y}sin\left(\theta_{s}\right)&\nonumber \\
&\pm2mcb\tilde{B}_{1y}sin\left(\theta_{s}\right)
-2imc\tilde{B}_{1x}sin\left(\theta_{s}\right))k_{z}
\times\frac{\partial f_{0}}{\partial p_{z}}\frac{e^{-ib\phi_{s}}}{\left(\omega-k_{z}p_{z}/m-b\omega_{cg}\right)}&\nonumber \\
&-\sum_{a=\pm1}\frac{q\mu}{4mc^{2}}\left(\frac{a}{2}\right)cos\theta_{s}(-iB_{0}\tilde{E}_{1x}
-\frac{2ic}{\mu}p_{z}\tilde{B}_{1y}-\frac{2ac}{\mu}p_{z}\tilde{B}_{1x}&\nonumber \\
&-\frac{2amc^{2}}{\mu}\tilde{E}_{1y}+\frac{2imc^{2}}{\mu}\tilde{E}_{1x})
\frac{\partial f_{0}}{\partial p_{\bot}}
\frac{e^{-ia\phi_{p}}}{\left(\omega-k_{z}p_{z}/m-a\omega_{c}\right)}&\nonumber \\
&-\sum_{a=\pm1}\frac{\mu}{4mc}(p_{\bot}\tilde{E}_{1y}cos\theta_{s}
+\frac{2aq}{\mu}p_{\bot}\tilde{B}_{1x}+\frac{2qi}{\mu}p_{\bot}\tilde{B}_{1y})k_{z}&\nonumber \\
&\times\frac{\partial f_{0}}{\partial p_{z}}
\frac{e^{-ia\phi_{p}}}{\left(\omega-k_{z}p_{z}/m-a\omega_{c}\right)}&\nonumber \\
&+\sum_{a=\pm1}\frac{\mu i}{4\hbar mc}p_{\bot}\left(1-a\right)\tilde{E}_{1z}\frac{\partial f_{0}}{\partial \theta_{s}}\frac{e^{-ia\phi_{p}}e^{-i\phi_{s}}}{\left(\omega-k_{z}p_{z}/m-a\omega_{c}-\omega_{cg}\right)}&\nonumber \\
&+\sum_{a=\pm1}\frac{\mu i}{4\hbar mc}p_{\bot}\left(1+a\right)\tilde{E}_{1z}\frac{\partial f_{0}}{\partial \theta_{s}}\frac{e^{-ia\phi_{p}}e^{i\phi_{s}}}{\left(\omega-k_{z}p_{z}/m-a\omega_{c}+\omega_{cg}\right)}&\nonumber \\
&-\sum_{a=\pm1}\frac{\mu qi}{8mc^{2}}B_{0}\tilde{E}_{1z}sin\theta_{s}\left(1-a\right)\frac{\partial f_{0}}{\partial p_{\bot}}\frac{e^{-ia\phi_{p}}e^{-i\phi_{s}}}{\left(\omega-k_{z}p_{z}/m-a\omega_{c}-\omega_{cg}\right)}&\nonumber \\
&+\sum_{a=\pm1}\frac{\mu qi}{8mc^{2}}B_{0}\tilde{E}_{1z}sin\theta_{s}\left(1-a\right)\frac{\partial f_{0}}{\partial p_{\bot}}\frac{e^{-ia\phi_{p}}e^{i\phi_{s}}}{\left(\omega-k_{z}p_{z}/m-a\omega_{c}+\omega_{cg}\right)}&\nonumber \\
&-\sum_{a=\pm1}\frac{\mu qi}{8mc^{2}}B_{0}\tilde{E}_{1z}sin\theta_{s}\left(1-a\right)\frac{\partial f_{0}}{\partial p_{\bot}}\frac{e^{-ia\phi_{p}}e^{-i\phi_{s}}}{\left(\omega-k_{z}p_{z}/m-a\omega_{c}-\omega_{cg}\right)}&\nonumber \\
&-\sum_{a=\pm1}\frac{\mu qi}{8mc^{2}}B_{0}\tilde{E}_{1z}cos\theta_{s}\left(1+a\right)\frac{\partial^{2} f_{0}}{\partial\theta_{s}\partial p_{\bot}}\frac{e^{-ia\phi_{p}}e^{i\phi_{s}}}{\left(\omega-k_{z}p_{z}/m-a\omega_{c}+\omega_{cg}\right)}&\nonumber \\
&-\sum_{a=\pm1}\frac{\mu qi}{8mc^{2}}B_{0}\tilde{E}_{1z}cos\theta_{s}\left(1-a\right)\frac{\partial^{2} f_{0}}{\partial\theta_{s}\partial p_{\bot}}\frac{e^{-ia\phi_{p}}e^{-i\phi_{s}}}{\left(\omega-k_{z}p_{z}/m-a\omega_{c}-\omega_{cg}\right)}&\nonumber \\
&-\sum_{a=\pm1}\frac{\mu i}{8mc}p_{\bot}\tilde{E}_{1z}k_{z}cos\theta_{s}\left(1+a\right)\frac{\partial^{2} f_{0}}{\partial\theta_{s}\partial p_{z}}\frac{e^{-ia\phi_{p}}e^{i\phi_{s}}}{\left(\omega-k_{z}p_{z}/m-a\omega_{c}
+\omega_{cg}\right)}&\nonumber \\
&+\sum_{a=\pm1}\frac{\mu i}{8mc}p_{\bot}\tilde{E}_{1z}k_{z}cos\theta_{s}\left(1-a\right)\frac{\partial^{2} f_{0}}{\partial\theta_{s}\partial p_{z}}\frac{e^{-ia\phi_{p}}e^{-i\phi_{s}}}{\left(\omega-k_{z}p_{z}/m-a\omega_{c}
-\omega_{cg}\right)}&\nonumber \\
&-\sum_{a=\pm1}\frac{\mu i}{8mc}p_{\bot}\tilde{E}_{1z}k_{z}sin\theta_{s}\left(1+a\right)\frac{\partial f_{0}}{\partial p_{z}}\frac{e^{-ia\phi_{p}}e^{i\phi_{s}}}{\left(\omega-k_{z}p_{z}/m-a\omega_{c}
+\omega_{cg}\right)}&\nonumber \\
&+\sum_{a=\pm1}\frac{\mu i}{8mc}p_{\bot}\tilde{E}_{1z}k_{z}sin\theta_{s}\left(1-a\right)\frac{\partial f_{0}}{\partial p_{z}}\frac{e^{-ia\phi_{p}}e^{-i\phi_{s}}}{\left(\omega-k_{z}p_{z}/m-a\omega_{c}
-\omega_{cg}\right)}&\nonumber \\
&+\sum_{a=\pm1}\frac{q\mu a}{4mc^{2}}B_{0}\tilde{E}_{1y}sin\theta_{s}\frac{\partial^{2}f_{0}}{\partial\theta_{s}\partial p_{\bot}}\frac{e^{-ia\phi_{p}}}{\left(\omega-k_{z}p_{z}/m-a\omega_{c}\right)}&\nonumber \\
&-\sum_{a=\pm1}\frac{q\mu i}{4mc^{2}}B_{0}\tilde{E}_{1x}sin\theta_{s}\frac{\partial^{2}f_{0}}{\partial\theta_{s}\partial p_{\bot}}\frac{e^{-ia\phi_{p}}}{\left(\omega-k_{z}p_{z}/m-a\omega_{c}\right)}&\nonumber \\
&-\sum_{a=\pm1}\frac{\mu ia}{4mc}p_{\bot}\tilde{E}_{1x}k_{z}sin\theta_{s}\frac{\partial^{2}f_{0}}{\partial\theta_{s}\partial p_{z}}\frac{e^{-ia\phi_{p}}}{\left(\omega-k_{z}p_{z}/m-a\omega_{c}\right)}&\nonumber \\
&+\sum_{a=\pm1}\frac{\mu }{4mc}p_{\bot}\tilde{E}_{1y}k_{z}sin\theta_{s}\frac{\partial^{2}f_{0}}{\partial\theta_{s}\partial p_{z}}\frac{e^{-ia\phi_{p}}}{\left(\omega-k_{z}p_{z}/m-a\omega_{c}\right)}&\nonumber \\
&-\mu \tilde{B}_{1z}k_{z}sin\theta_{s}\frac{\partial^{2}f_{0}}{\partial\theta_{s}\partial p_{z}}\frac{1}{\left(\omega-k_{z}p_{z}/m\right)}&\nonumber \\
&+(\mu \tilde{B}_{1z}k_{z}cos\theta_{s}
-iq\tilde{E}_{1z})\frac{\partial f_{0}}{\partial p_{z}}\frac{1}{\left(\omega-k_{z}p_{z}/m\right)}&
\end{eqnarray}
here $\omega_{c}=-\frac{eB_{0}}{mc}$ and $\omega_{cg}=-\frac{2\mu
B_{0}}{\hbar}$ are the electron cyclotron frequency and the spin
precession frequency, respectively.

Derivation of the general dispersion relation, in the same way as in
the classical cases, is based on the relation $det D_{ij}=0$ with
$D_{ij}=\delta_{ij}\left(1-\frac{c^{2}k^{2}}{\omega^{2}}\right)
-\frac{k_{i}k_{j}c^{2}}{\omega^{2}}\delta_{ij}+i\frac{\sigma_{ij}}{\varepsilon_{0}\omega}$,
where the quantities $\delta_{ij}$, $\varepsilon_{0}$ and
$\sigma_{ij}$ are the Kronecker delta function, the permitivity
constant and the conductivity tensor, respectively. The conductivity
tensor is proportional to the current density and the electric field
as $J_{i}=\sum_{j}\sigma_{ij}E_{j}$. Here, the total current density
includes three different contributions so that;
\begin{equation}
  J=J_{F}+\nabla\times M+\frac{\partial P}{\partial t}
\end{equation}
where, the first term is introducing the free current density, the
second and last ones are introducing the magnetization, $M$, and the
polarization, $P$, contribution, respectively, due to the spin which
are defined respectively as:
\begin{equation}
  J_{F}=q\int d\Omega\left(\frac{p}{m}+\frac{3\mu}{2mc}E\times s\right)f
\end{equation}
\begin{equation}
  M=3\mu\int d\Omega sf
\end{equation}
and
\begin{equation}
  P=-3\mu\int d\Omega \frac{s\times p}{2mc}f
\end{equation}
Here, $d \Omega=d^{3}vd^{2}s$ is the integration measure performed
over the three velocity variable in the cylindrical space (which
will be transformed to the momentum variations for simplicity) and
the two spin degrees of freedom in the spherical space. Notice that,
the current density is only included the electron contribution and
the ions contributions are ignored because of the their larger mass
and intrinsic magnetic moment lower than the electrons one. Now, let
us restrict the results to the specific situation. we consider the
polarization as $\textbf{E}_{1} = E_{1y} \hat{y}$ so that
$\textbf{B}_{1} = B_{1x}\hat{x}$. This polarization can be justified
only when $\sigma_{yy}\gg\sigma_{xy}, \sigma_{zy}$, so the
dispersion relation can be presented (by the Ampere’s law) as
follows:
\begin{equation}
  \omega^{2}-c^{2}k_{z}^{2}+\frac{i\omega}{\varepsilon_{0}}\sigma_{yy}=0
\end{equation}
where
\begin{eqnarray}
&\sigma_{yy}=+\sum_{\nu=+,-}\nu\frac{q^{2}\mu i}{16m^{2}c^{2}}\int p_{\bot}B_{0}\frac{\partial F_{0\nu}}{\partial p_{\bot}}\frac{d^{3}p}{\left(\omega-k_{z}p_{z}/m-\omega_{c}\right)}&\nonumber\\
&+\sum_{\nu=+,-}\nu\frac{q\mu i}{8m^{2}c}\int p_{\bot}^{2}k_{z}\frac{\partial F_{0\nu}}{\partial p_{z}}\frac{d^{3}p}{\left(\omega-k_{z}p_{z}/m-\omega_{c}\right)}&\nonumber\\
&-\sum_{\nu=+,-}\frac{q^{2}i}{4m^{2}}\int p_{\bot}^{2}\frac{k_{z}}{\omega}\frac{\partial F_{0\nu}}{\partial p_{z}}\frac{d^{3}p}{\left(\omega-k_{z}p_{z}/m-\omega_{c}\right)}&\nonumber\\
&-\sum_{\nu=+,-}\frac{q^{2}i}{4m\omega}\int p_{\bot}\frac{\left(\omega-k_{z}p_{z}/m\right)}{\left(\omega-k_{z}p_{z}/m-\omega_{c}\right)}\frac{\partial F_{0\nu}}{\partial p_{\bot}}d^{3}p&\nonumber\\
&+\sum_{\nu=+,-}\sum_{b=\pm1}\frac{3i\mu^{2}b\nu}{10m\hbar}\int p_{z}k_{z}F_{0\nu}\frac{d^{3}p}{\left(\omega-k_{z}p_{z}/m-b\omega_{cg}\right)}&\nonumber\\
&+\sum_{\nu=+,-}\sum_{b=\pm1}\frac{i\mu^{2}b\nu}{\hbar}\int \frac{c^{2}k_{z}^{2}}{\omega}F_{0\nu}\frac{d^{3}p}{\left(\omega-k_{z}p_{z}/m-b\omega_{cg}\right)}&\nonumber\\
&-\sum_{\nu=+,-}\sum_{b=\pm1}\frac{i\mu^{2}}{2}\int \frac{c^{2}k_{z}^{3}}{\omega}\frac{\partial F_{0\nu}}{\partial p_{z}}\frac{d^{3}p}{\left(\omega-k_{z}p_{z}/m-b\omega_{cg}\right)}&\nonumber\\
&-\sum_{\nu=+,-}\sum_{b=\pm1}\frac{11i\omega\mu^{2}b\nu}{20m^{2}c^{2}\hbar}\int p_{z}^{2}F_{0\nu}\frac{d^{3}p}{\left(\omega-k_{z}p_{z}/m-b\omega_{cg}\right)}&\nonumber\\
&-\sum_{\nu=+,-}\sum_{b=\pm1}\frac{i\mu^{2}b\nu c}{2mc\hbar}\int p_{z}k_{z}F_{0\nu}\frac{d^{3}p}{\left(\omega-k_{z}p_{z}/m-b\omega_{cg}\right)}&\nonumber\\
&-\sum_{\nu=+,-}\sum_{b=\pm1}\frac{i\mu^{2}\omega}{8m^{2}c^{2}}\int p_{z}^{2}k_{z}\frac{\partial F_{0\nu}}{\partial p_{z}}\frac{d^{3}p}{\left(\omega-k_{z}p_{z}/m-b\omega_{cg}\right)}&\nonumber\\
&+\sum_{\nu=+,-}\sum_{b=\pm1}\frac{i\mu^{2}}{4m}\int p_{z}k_{z}^{2}\frac{\partial F_{0\nu}}{\partial p_{z}}\frac{d^{3}p}{\left(\omega-k_{z}p_{z}/m-b\omega_{cg}\right)}&\nonumber\\
&-\sum_{\nu=+,-}\sum_{a=\pm1}\frac{iq\mu^{2}a\omega}{32m^{2}c^{3}}\int p_{\bot}B_{0}\frac{\partial F_{0\nu}}{\partial p_{\bot}}\frac{d^{3}p}{\left(\omega-k_{z}p_{z}/m-a\omega_{c}\right)}&\nonumber\\
&-\sum_{\nu=+,-}\sum_{a=\pm1}\frac{i\mu^{2}\omega}{16m^{2}c^{2}}\int p_{\bot}^{2}k_{z}\frac{\partial F_{0\nu}}{\partial p_{z}}\frac{d^{3}p}{\left(\omega-k_{z}p_{z}/m-a\omega_{c}\right)}&\nonumber\\
&-\sum_{\nu=+,-}\sum_{a=\pm1}\frac{i\mu aq\nu}{8m^{2}c}\int p_{\bot}p_{z}k_{z}\frac{\partial F_{0\nu}}{\partial p_{\bot}}\frac{d^{3}p}{\left(\omega-k_{z}p_{z}/m-a\omega_{c}\right)}&\nonumber\\
&+\sum_{\nu=+,-}\sum_{a=\pm1}\frac{i\mu aq\nu}{8m^{2}c}\int p_{\bot}^{2}k_{z}\frac{\partial F_{0\nu}}{\partial p_{z}}\frac{d^{3}p}{\left(\omega-k_{z}p_{z}/m-a\omega_{c}\right)}&\nonumber\\
&+\sum_{\nu=+,-}\sum_{a=\pm1}\frac{i\omega\mu aq\nu}{8mc}\int p_{\bot}\frac{\partial F_{0\nu}}{\partial p_{\bot}}\frac{d^{3}p}{\left(\omega-k_{z}p_{z}/m-a\omega_{c}\right)}&\nonumber\\
&+\sum_{\nu=+,-}\nu\frac{q^{2}\mu i}{16m^{2}c^{2}}\int p_{\bot}B_{0}\frac{\partial F_{0\nu}}{\partial p_{\bot}}\frac{d^{3}p}{\left(\omega-k_{z}p_{z}/m+\omega_{c}\right)}&\nonumber\\
&-\sum_{\nu=+,-}\nu\frac{q\mu i}{8m^{2}c}\int p_{\bot}^{2}k_{z}\frac{\partial F_{0\nu}}{\partial p_{z}}\frac{d^{3}p}{\left(\omega-k_{z}p_{z}/m+\omega_{c}\right)}&\nonumber\\
&- \sum_{\nu=+,-}\frac{q^{2}i}{4m^{2}}\int p_{\bot}^{2}\frac{k_{z}}{\omega}\frac{\partial F_{0\nu}}{\partial p_{z}}\frac{d^{3}p}{\left(\omega-k_{z}p_{z}/m+\omega_{c}\right)}&\nonumber\\
&-\sum_{\nu=+,-}\frac{q^{2}i}{4m\omega}\int p_{\bot}\frac{\left(\omega-k_{z}p_{z}/m\right)}{\left(\omega-k_{z}p_{z}/m+\omega_{c}\right)}\frac{\partial F_{0\nu}}{\partial p_{\bot}}d^{3}p&
\end{eqnarray}
In the following, defining the exact form of the distribution
function can represent an analytical form of the dispersion
relation. Our purpose is investigating the electromagnetic
instabilities excited by the temperature anisotropy, therefore it
can be defined:
\begin{eqnarray}
&F_{0\nu}=\left[\frac{n_{0}}{\left(2\pi K_{B}\right)^{\frac{3}{2}}m^{\frac{3}{2}}T_{\bot}T_{z}^{\frac{1}{2}}}\right]  \exp\left[-\left(\frac{p_{\bot}^{2}}{2m K_{B}T_{\bot}}+\frac{p_{z}^{2}}{2m K_{B}T_{z}}\right)\right]&\nonumber\\
&\times\left(\frac{e^{\nu\mu B_{0}/K_{B}T_{sp}}}{e^{\mu B_{0}/K_{B}T_{sp}}+e^{-\mu B_{0}/K_{B}T_{sp}}}\right)&
\end{eqnarray}
The quantity $T_{sp}$ is defined as the spin temperature [14]. In
addition to the free energy of the velocity space, the plasma can be
confronted with another free energy, which is supposed to be the
difference between the high- and low-energy spin. In that case, the
number of particles in the two spin states does not correspond to
thermodynamic equilibrium. As we know, we have two instabilities,
namely spin instability and velocity space instability. The first
one comes from the spin temperature, $T_{sp}$, and the second one
comes from the kinetic energy; the two together form the source of
such instability. In fact, it is expected that spin instabilities
exist together with velocity space instabilities when the deviation
in the spin temperature from the kinetic anisotropic temperature can
be the source of this instability. Notice that when the only degree
of particle freedom is spin, and all particles have the lowest
energy (down spin state), the entropy will be equal to zero. In this
case, adding the energy and flip-up particle spin leads to increase
in the entropy and positive temperature until reaching the maximum
entropy, where one-half of particles have the down spin state. After
that, increasing the particle number in the up spin state leads to a
decrease in the entropy and temperature, so that when all particles
have the up spin state, the entropy is equal to zero and the
temperature is negative. In the presence of other degrees of freedom
(here, kinetic degrees relevant to the velocity space), the
condition is special. In this situation, when the spin variations
are independent of the other freedom variables, the definition of
spin temperature can be important. In this condition, in the
presence of a strong external magnetic field, the coupling between
the degrees of freedom of spin and velocity will be sufficiently
weak, while the coupling between spin freedoms is strong; therefore
the timescale of energy flow between the degrees of freedom is large
and the spin temperature will be negative. Here it is assumed that
the energy difference between the high- and low-energy spin states
is small, but even this small value can be important for generating
the free energy in the background magnetic field [14].

Now, replacing the above distribution in the Eq. (12), the dispersion relation can be derived as:
\begin{eqnarray}
&\omega^{2}-c^{2}k_{z}^{2}-\omega_{p}^{2}(1-\frac{T_{\bot}}{T_{z}})
+\frac{g^{2}\hbar^{2}k_{z}^{2}}{32mK_{B}T_{z}}\omega_{p}^{2}(2+\zeta_{2}Z(\zeta_{2})+\zeta_{1}Z(\zeta_{1}))&\nonumber\\
&-\frac{\omega\mu}{8\sqrt{2}m^{1/2}c^{2}}B_{0}\omega_{p}^{2}\frac{tanh\alpha}{K_{B}^{1/2}T_{z}^{1/2}k_{z}}
[Z(\zeta_{4})+Z(\zeta_{3})]+\frac{\omega_{p}^{2}}{2}\frac{T_{\bot}}{T_{z}}[\zeta_{4}Z(\zeta_{4})+\zeta_{3}Z(\zeta_{3})]&\nonumber\\
&-\frac{\omega g\hbar}{16mc^{2}}\omega_{p}^{2}\frac{T_{\bot}}{T_{z}}tanh\alpha[\zeta_{4}Z(\zeta_{4})-\zeta_{3}Z(\zeta_{3})]
+\frac{\omega_{p}^{2}}{2}\omega_{c}\frac{\sqrt{m}}{\sqrt{2K_{B}T_{z}}k_{z}}[Z(\zeta_{4})-Z(\zeta_{3})]&\nonumber\\
&+\frac{3\omega}{160}\frac{g^{2}\hbar}{mc^{2}}\omega_{p}^{2}tanh\alpha[\zeta_{2}Z(\zeta_{2})-\zeta_{1}Z(\zeta_{1})]
+\frac{g^{2}\hbar k_{z}}{16\sqrt{2mK_{B}T_{z}}}\omega_{p}^{2}tanh\alpha[Z(\zeta_{2})-Z(\zeta_{1})]&\nonumber\\
&-\frac{24\omega^{2}g^{2}\hbar }{320m^{2}c^{4}k_{z}}\omega_{p}^{2}tanh\alpha\sqrt{2mK_{B}T_{z}}[\zeta_{2}^{2}Z(\zeta_{2})
-\zeta_{1}^{2}Z(\zeta_{1})+\zeta_{2}-\zeta_{1}]&\nonumber\\
&-\frac{\omega g^{2}\hbar }{320mc^{4}}\omega_{p}^{2}tanh\alpha[\zeta_{2}Z(\zeta_{2})-\zeta_{1}Z(\zeta_{1})]
+\frac{\omega^{2}g^{2}\hbar^{2}}{64m^{2}c^{4}}\omega_{p}^{2}
[\zeta_{2}^{3}z(\zeta_{2})-\zeta_{1}^{3}z(\zeta_{1})+\zeta_{2}^{2}]&\nonumber\\
&-\frac{\omega g^{2}\hbar^{2}k_{z}}{32m^{3/2}c^{2}\sqrt{2K_{B}T_{z}}}\omega_{p}^{2}
[\zeta_{2}^{2}z(\zeta_{2})+\zeta_{1}^{2}z(\zeta_{1})-\zeta_{2}-\zeta_{1}]&\nonumber\\
&-\frac{\omega^{2} g\hbar^{2}\omega_{cg}}{128m^{3/2}c^{4}\sqrt{2K_{B}T_{z}}}\frac{\omega_{p}^{2}}{k_{z}}
[z(\zeta_{4})-z(\zeta_{3})]
+\frac{\omega^{2}g^{2}\hbar^{2}}{128m^{2}c^{4}}\omega_{p}^{2}\frac{T_{\bot}}{T_{z}}
[\zeta_{4}z(\zeta_{4})+\zeta_{3}z(\zeta_{3})]&\nonumber\\
&+\frac{\omega^{2}}{64}\frac{g^{2}\hbar^{2}}{m^{2}c^{4}}\omega_{p}^{2}\frac{T_{\bot}}{T_{z}}
-\frac{\omega}{16}\frac{g\hbar}{mc^{2}}\omega_{p}^{2}(\frac{T_{\bot}}{T_{z}}-1)
[\zeta_{4}Z(\zeta_{4})-\zeta_{3}Z(\zeta_{3})]&\nonumber\\
&-\frac{\omega^{2}}{16}\frac{g\hbar}{k_{z}c^{2}}\frac{\omega_{p}^{2}}{\sqrt{2mK_{B}T_{z}}}[Z(\zeta_{4})-Z(\zeta_{3})]=0&
\end{eqnarray}
The function $Z(\zeta)$ is the plasma dispersion function [15] given
by
$Z(\zeta)=\frac{1}{\sqrt{\pi}}\int_{-\infty}^{+\infty}\frac{e^{-x^{2}}}{x-\zeta}dx$
with the arguments
$\zeta_{1}=\frac{\sqrt{m}}{\sqrt{2K_{B}T_{z}}k_{z}}(\omega+\omega_{cg})$,
$\zeta_{2}=\frac{\sqrt{m}}{\sqrt{2K_{B}T_{z}}k_{z}}(\omega-\omega_{cg})$,
$\zeta_{3}=\frac{\sqrt{m}}{\sqrt{2K_{B}T_{z}}k_{z}}(\omega+\omega_{c})$,
$\zeta_{4}=\frac{\sqrt{m}}{\sqrt{2K_{B}T_{z}}k_{z}}(\omega-\omega_{cg})$
and $x=\frac{p_{z}}{\sqrt{2mK_{B}T_{z}}}$. Here, let us follow the
results to the non resonant electromagnetic instabilities. In this
order, it is necessary, the arguments of the plasma dispersion
function be comparable to or smaller than unity [16]. Therefore, we
consider, two specific limiting conditions: the arguments of the
plasma dispersion function higher than unity and the arguments
smaller than unity. In the limit of arguments larger than the unity,
the function $Z(\zeta)$ can be approximated by $Z(\zeta)
=-\frac{1}{\zeta}-\frac{1}{\zeta^{3}}+\cdots$, where the dispersion
relation will be a fully real relation that cannot produce a
simplified form of the real frequency and the growth rate of the
instability. In the opposite condition, the plasma dispersion
function can be approximated as
$Z(\zeta)=-2\zeta+\cdots+i\sqrt{\pi}$, so that it is obtained (for
the sub-luminal waves):
\begin{eqnarray}
&c^{2}k_{z}^{2}+\omega_{p}^{2}-i\sqrt{\pi}\frac{\omega}{8}\frac{g\hbar}{mc^{2}}
\omega_{c}\omega_{p}^{2}\frac{\sqrt{m}}{\sqrt{2K_{B}T_{z}}k_{z}}(\frac{T_{\bot}}{T_{z}}-1)
-\omega_{p}^{2}\frac{T_{\bot}}{T_{z}}&\nonumber\\
&-i\sqrt{\pi}\frac{\omega}{40}\frac{g^{2}\hbar}{mc^{2}}
\omega_{cg}\omega_{p}^{2}\frac{\sqrt{m}}{\sqrt{2K_{B}T_{z}}k_{z}}tanh\alpha
-i\sqrt{\pi}\frac{g^{2}\hbar^{2}k_{z}}{16\sqrt{2mK_{B}T_{z}}}\frac{\omega}{K_{B}T_{z}}\omega_{p}^{2}&\nonumber\\
&-\frac{g^{2}\hbar^{2}k_{z}^{2}}{16mK_{B}T_{z}}\omega_{p}^{2}
-\frac{g^{2}\hbar}{8K_{B}T_{z}}\omega_{p}^{2}\omega_{cg}tanh\alpha
-i\sqrt{\pi}\frac{T_{\bot}}{T_{z}}\frac{\sqrt{m}}{\sqrt{2K_{B}T_{z}}k_{z}}\omega_{p}^{2}\omega&\nonumber \\
&-i\sqrt{\pi}\frac{T_{\bot}}{T_{z}}\frac{\omega}{8\sqrt{2mK_{B}T_{z}}k_{z}c^{2}}\omega_{p}^{2}\omega_{c}tanh\alpha
+2i\sqrt{\pi}\frac{\omega}{8\sqrt{2mK_{B}T_{z}}k_{z}c^{2}}\omega_{p}^{2}\mu B_{0}tanh\alpha
=0&
\end{eqnarray}
The wave frequency, $\omega$, is sum of the real and imaginary parts
which define the frequency and the growth rate of the instability as
$\omega_{r}$ and $\omega_{i}$, respectively. According this, it is
obtained:
\begin{equation}
  \omega_{r}=0
\end{equation}
and
\begin{eqnarray}
  &\omega_{i}=\frac{v_{th,z}k_{z}}{\sqrt{\pi}}[1-\frac{(1+\frac{c^{2}k_{z}^{2}}{\omega_{p}^{2}})}{\frac{T_{\bot}}{T_{z}}}
  +\frac{g^{2}\hbar^{2}k_{z}^{2}}{16mK_{B}T_{\bot}}(1+2\frac{m}{\hbar k_{z}^{2}}\omega_{cg}tanh\alpha)]&\nonumber\\
  &\times[1+\frac{\hbar}{4mc^{2}}\omega_{cg}[1-\frac{T_{z}}{T_{\bot}}(1+\frac{tanh\alpha}{2})+tanh\alpha]
  +\frac{T_{z}}{T_{\bot}}\frac{g^{2}\hbar}{40mc^{2}}\omega_{cg}tanh\alpha
  +\frac{g^{2}\hbar^{2}k_{z}^{2}}{16mK_{B}T_{\bot}}]^{-1}&
\end{eqnarray}
In the absence of the spin relativistic effects,
$\frac{\hbar\omega_{cg}}{mc^{2}}\rightarrow0$,the results goes to
the previous results where the quantum effects due to the electrons
spin were investigated in a fully non relativistic plasmas [14]. For
the non resonant instabilities, it is necessary the instability
growth rate be larger than the instability frequency so that the
growth condition of the instability is satisfied as follows:
  \begin{equation}
  k_{z}<k_{cut off}=\left[1-\frac{T_{z}}{T_{\bot}}+\frac{g^{2}\hbar}
  {8K_{B}T_{\bot}}\omega_{cg}\tanh\left(\frac{\mu_{e}B_{0}}{K_{B}T_{sp}}\right)\right]^{\frac{1}{2}}
  \left[\frac{c^{2}}{\omega_{pe}^{2}}\frac{T_{z}}{T_{\bot}}
  -\frac{g^{2}\hbar^{2}}{16m_{e}K_{B}T_{\bot}}\right]^{-\frac{1}{2}}
\end{equation}
where, the instability does not have any fluctuation in time and the
results are similar to the non relativistic cases [14].
\section{Physical systems}
The astrophysical and laboratory plasmas in particular inertial
confinement fusion (ICF) are always two of the interesting plasma
systems for studying the electromagnetic instabilities due to the
temperature anisotropy, where these can play important roles because
of generating strong magnetic fields [17-19]. It is well known,
these instabilities can be excited because of all laser-plasma
energy deposition processes and as a result of the propagation of a
shock wave in the first and second ones, respectively. In other
hand, it is well known, the quantum effects can start playing a role
in the high-density regime where, such dense relativistic plasmas
can be observed in particular in stars and planetary interiors for
astrophysical subjects and the transmission of the corona of the
fuel plasma to the core of fuel pellet in ICF plasmas [20, 21].
Therefore, in the following work, the obtained results in the last
section are investigated for the ICF and astrophysical subjects
respectively.

\subsection{Inertial confinement fusion plasmas (ICF)}
The obtained growth rate (Eq. (17)) includes different parameters
every one of which can play a significant role in exact numerical
investigations, for example the temperature anisotropy fraction
$\frac{T_{\bot}}{T_{z}}$, the number density relevant to the plasma
frequency, the field intensity relevant to the electron cyclotron
and the spin precession frequency. The results show that, otherwise
it is expected that the new effects lead to new findings but there
will not be significant difference in values of the normalized
growth rate compared to the last results in the fully non
relativistic spin polarized cases [14] and even to the classical
cases [22] because of the weak magnetic field (see the Fig. (1)).
Here, increasing the strength of the magnetic field (equivalent to
compression the cyclotron movement) leads to decreasing the growth
rate of the instability where this can not be high because of the
small contribution of the spin effects. The variation of the
normalized growth rate is shown in Fig. (2) for two different
strengths of the magnetic field $8 T$ and $25000 T$ and the fixed
$T_{z}=5000 eV$, $\frac{T_{\bot}}{T_{Z}}=5$, $n_{0}=10^{30} m^{-3}$
and $\frac{T_{sp}}{T_{z}}=2$. The contribution of the sentences
dependent on the weak relativistic effects can be affected by the
variations of the temperature anisotropy fraction
$\frac{T_{\bot}}{T_{z}}$, too. The illustration, the Fig. (3), shows
that decreasing this (equivalent to decreasing the free energy) can
lead to decreasing the maximum normalized growth rate about $61.6\%$
for example by decreasing the temperature anisotropy fraction about
two units. In other words, variations of the the electron number
density can be important for investigating the variations of the
normalized growth rate. Here, it is observed, that decreasing the
electron number density about 10 times can decrease the normalized
growth rate about $68.75\%$( Fig. (4). Notice that, all of these
variations can be equal to the obtained results for the non
relativistic spin polarized cases by ignorable values in similar
situation.
\subsection{Astrophysical plasmas}
Variations of the normalized growth rate are investigated for the
astrophysical plasmas, too. It was observed, that the weak
relativistic effects due to the spin will increase the instability
growth rate compared to that our previous non relativistic work [14]
(see the Fig. (5)). In fact, it is expected, the relativistic
effects due to the spin, lead to increasing the free energy in the
system and decreasing the electrons dissipation energy. Increasing
of the instability growth rate can be affected by strength of the
magnetic field. The results imply that, increasing the strength of
the magnetic field about 10 times leads to decreasing the maximum
normalized growth rate about $1.26\%$. Fig. (6) is illustrating the
curve variations of the normalized growth rate for the strength of
the magnetic field equal to $10^{7} T$ and $10^{8} T$ in the fixed
$\frac{T_{\bot}}{T_{z}}=5$, $n_{0}=10^{32} m^{-3}$,
$\frac{T_{sp}}{T_{z}}=2$ and $T_{z}=20000 eV$. The temperature
anisotropy fraction can be important, too, where it can play a role
in the sentences relevant to the weak relativistic effects. It is
expected, that the normalized growth rate decrease about $63.4\%$ by
decreasing the values of the temperature anisotropy fraction about 2
unit (Fig.(7)). Notice that, here, increasing the temperature
anisotropy fraction and strength of the magnetic field can lead to
excited relativistic effects and the quantization Landau effects
which are ignored here. Finally, the results are investigated for
the variation of the electron number density too. Fig. (8) shows,
that decreasing the electron number density about 10 times can lead
to decreasing the maximum value of the normalized growth rate about
$68.79\%$.
\section{Conclusion}
In this paper, the weak relativistic effects due to the electrons
spin are investigated on the electromagnetic instabilities excited
by the temperature anisotropy for two real physical situations ie.
the inertial confinement fusion and the astrophysical plasma.
Calculation model is based on the kinetic theory. This order, the
Foldy-Wouthuysen transformation for introducing the particle
Hamiltonian in external fields and the the Wigner transformation in
the phase space and the Q-transformation in the spin space were
applied for the evolution equation at the spatial scale higher than
the thermal de-Broglie length . The intrinsic relativistic effects
(such as the relativistic mass correction) have been ignored and
only the weak relativistic effects due to the electrons spin are
considered where the only presented quantum effects are relevant to
the spin. The results imply that, similar to our previous results,
the weak relativistic spin polarized effect can not be effective on
the electromagnetic instabilities in the ICF plasmas because of the
low strength of the magnetic field while this is different for the
astrophysical subjects. In the astrophysical plasmas, the weak
relativistic effect due to the electrons spin leads to increasing
the instability growth rate where the condition governed on growth
of the instability is unchanged (compared to the non relativistic
cases). In fact, here, it is expected, that the weak relativistic
effect due to the electron spin lead to increasing the free energy
in plasma so that the particles will be able to transmit in the
magnetic field more simply. \clearpage


\clearpage
\begin{figure}
\centerline{ \epsfxsize=10cm \epsffile{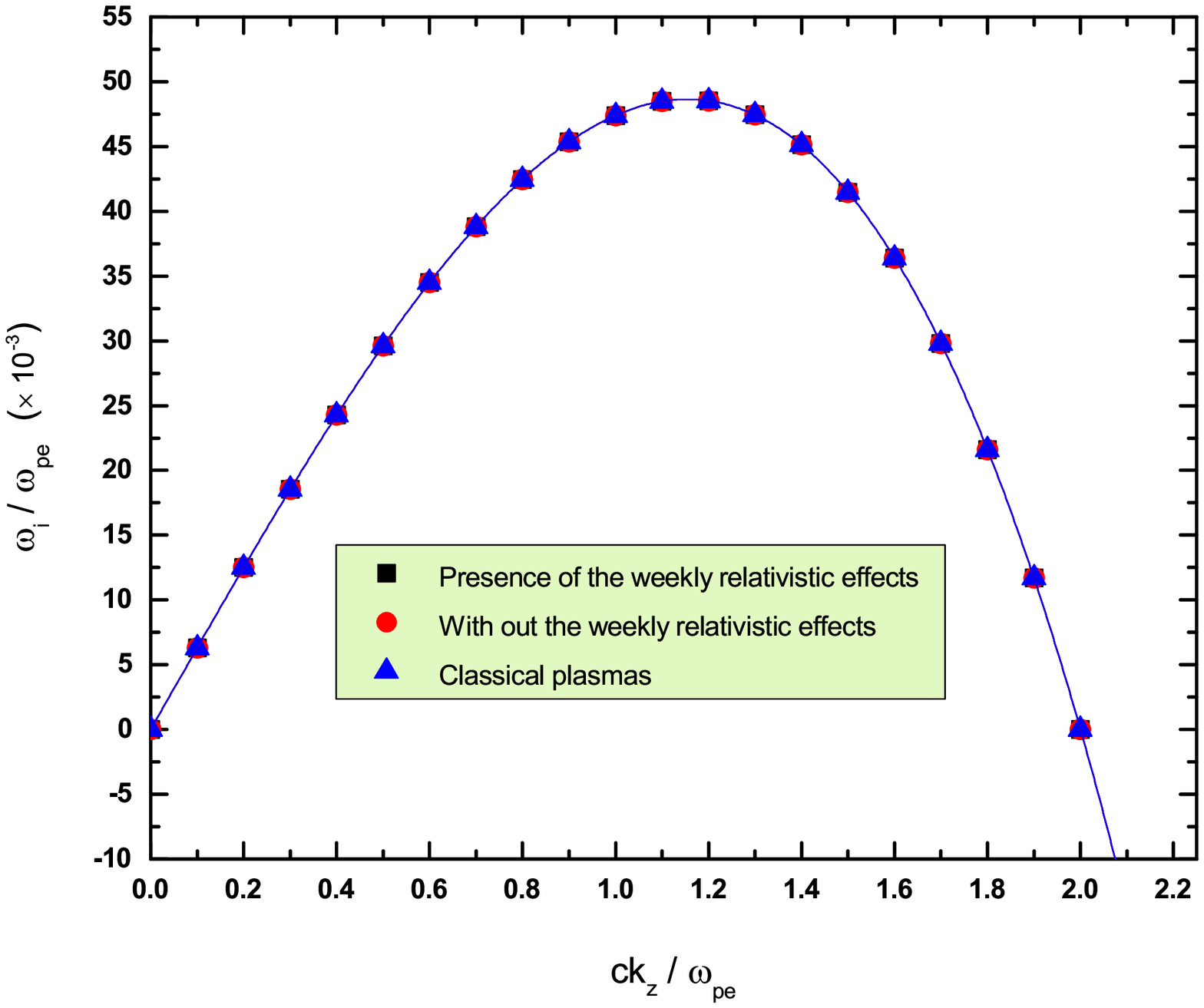}} \caption {The
normalized growth rate of the instability,
$\frac{\omega_{i}}{\omega_{pe}}$ as a function of the wave number,
$\frac{ck_{z}}{\omega_{pe}}$, for the non relativistic spin, the
classical and the weekly relativistic spin ICF plasmas in the fixed $T_{z}=5000 eV$, $n_{0}=10^{30}
m^{-3}$, $\frac{T_{\bot}}{T_{z}}=5$, $B_{0}=8 T$ and $\frac{T_{sp}}{T_{z}}=2$.}
\end{figure}
\clearpage
\begin{figure}
\centerline{ \epsfxsize=10cm \epsffile{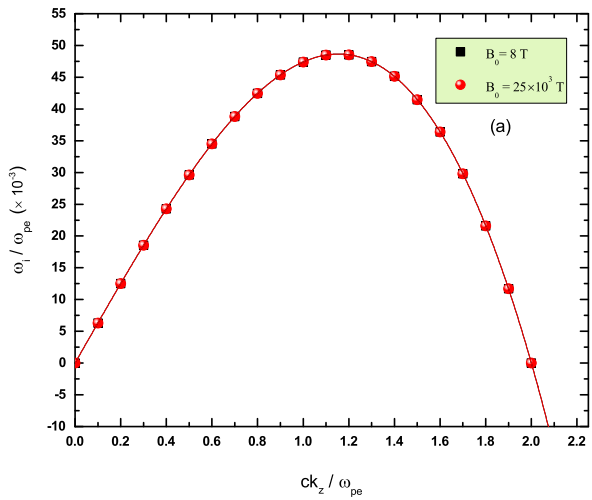}\vspace{1cm}
\epsfxsize=10cm\epsffile{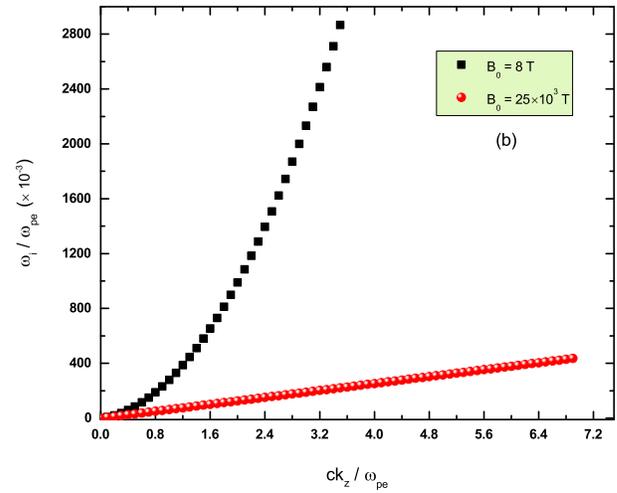}} \caption {a)The normalized growth rate
of the instability, $\frac{\omega_{i}}{\omega_{pe}}$, as a function of the wave number,
$\frac{ck_{z}}{\omega_{pe}}$ for different values of magnetic field
in the fixed $T_{z}=5000 eV$, $\frac{T_{sp}}{T_{z}}=2$ and
$n_{0}=10^{30} m^{-3}$. b)The normalized growth rate is illustrated
only for the sentences including spin effects.}
\end{figure}
\clearpage
\begin{figure}
\centerline{ \epsfxsize=10cm \epsffile{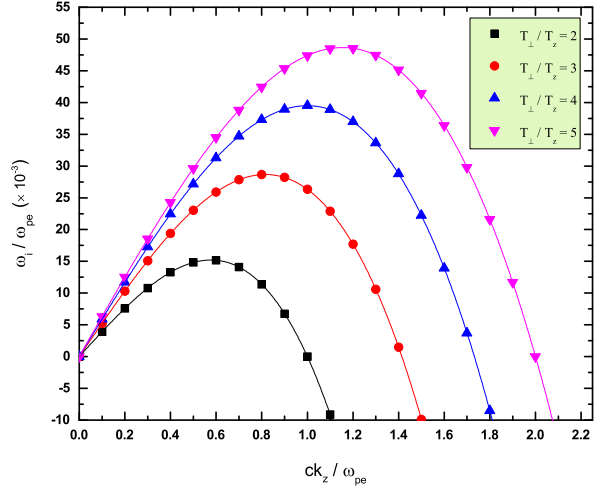}} \caption {The
normalized growth rate of the instability,
$\frac{\omega_{i}}{\omega_{pe}}$ as a function of the wave number,
$\frac{ck_{z}}{\omega_{pe}}$ for different values of the temperature
anisotropy fraction $\frac{T_{\bot}}{T_{z}}$ in the fixed $\frac{T_{sp}}{T_{z}}=2$,
$n_{0}=10^{30} m^{-3}$, $T_{z}=5000 eV$ and the magnetic
field $8 T$ for ICF subjects.}
\end{figure}
\clearpage
\begin{figure}
\centerline{ \epsfxsize=10cm \epsffile{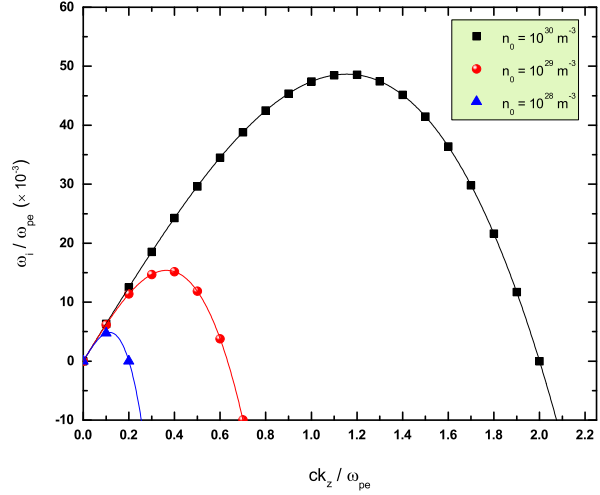}} \caption {The
normalized growth rate of the instability,
$\frac{\omega_{i}}{\omega_{pe}}$ as a function of the wave number,
$\frac{ck_{z}}{\omega_{pe}}$ for different values of the electron number density
in the fixed $\frac{T_{sp}}{T_{z}}=2$,
$\frac{T_{\bot}}{T_{z}}=5$, $T_{z}=5000 eV$ and the magnetic
field equal to $8 T$.}
\end{figure}
\clearpage
\begin{figure}
\centerline{ \epsfxsize=10cm \epsffile{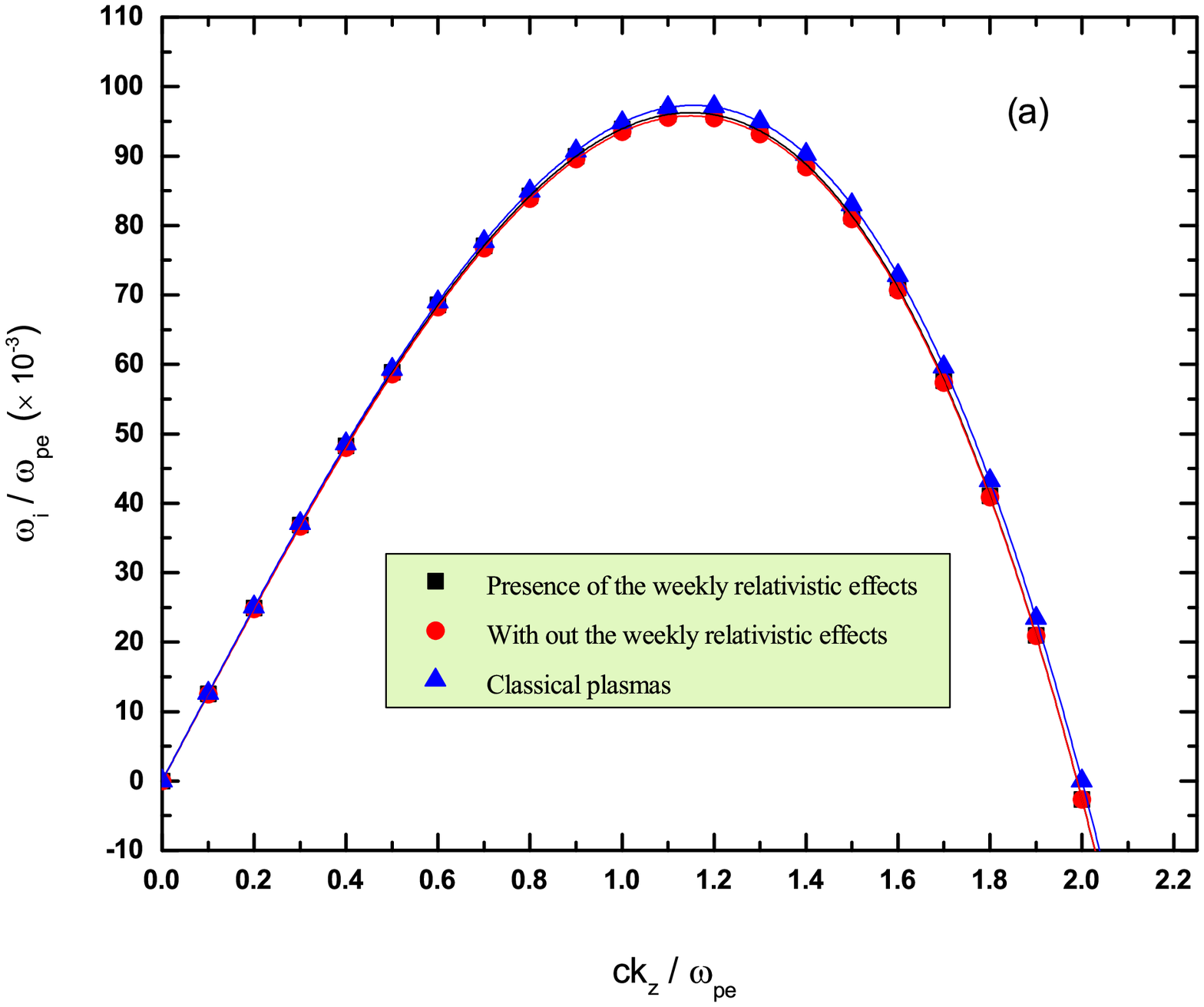}\vspace{1cm}
\epsfxsize=10cm\epsffile{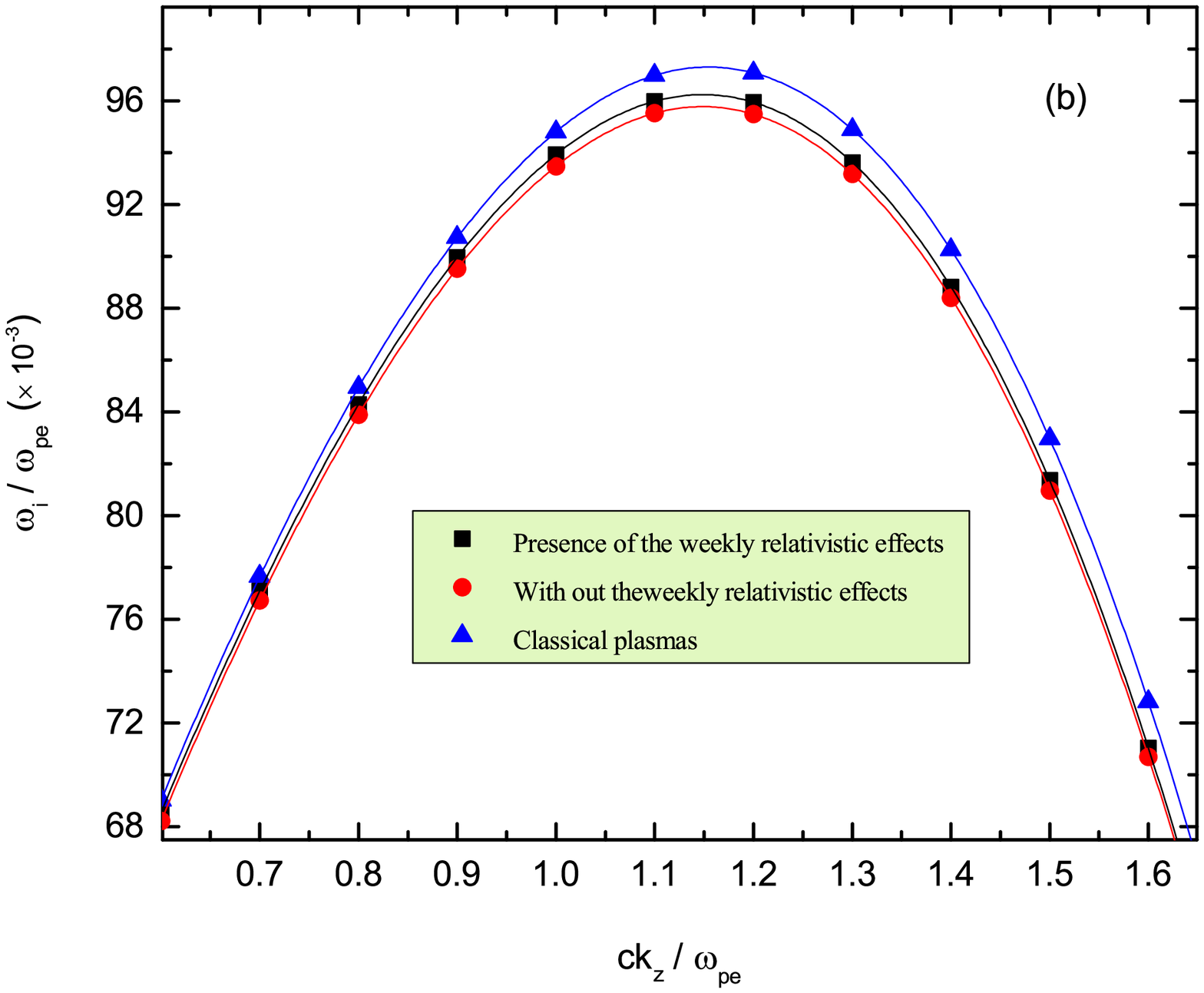}} \caption {a) The normalized
growth rate of the instability, $\frac{\omega_{i}}{\omega_{pe}}$ as a
function of the wave number, $\frac{ck_{z}}{\omega_{pe}}$, for the non relativistic spin, the classical and the weekly relativistic spin astrophysical plasmas in the fixed
$T_{z}=20000 eV$, $n_{0}=10^{32} m^{-3}$, $\frac{T_{\bot}}{T_{z}}=5$, $B_{0}=10^{8} T$
and $\frac{T_{sp}}{T_{z}}=2$. b) The variations of the normalizes growth rate in the part
(a) at the larger scale.}
\end{figure}
\clearpage
\begin{figure}
\centerline{ \epsfxsize=10cm \epsffile{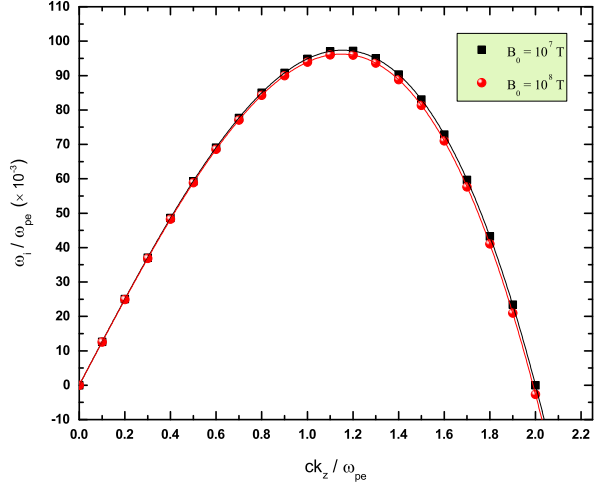}}\caption {The normalized
growth rate of the instability, $\frac{\omega_{i}}{\omega_{pe}}$ as a
function of the wave number, $\frac{ck_{z}}{\omega_{pe}}$ in the fixed
$T_{z}=20000 eV$, $n_{0}=10^{32} m^{-3}$, $\frac{T_{\bot}}{T_{z}}=5$
and $\frac{T_{sp}}{T_{z}}=2$ for different values of the magnetic
field in the astrophysical subjects.}
\end{figure}
\clearpage
\begin{figure}
\centerline{ \epsfxsize=10cm \epsffile{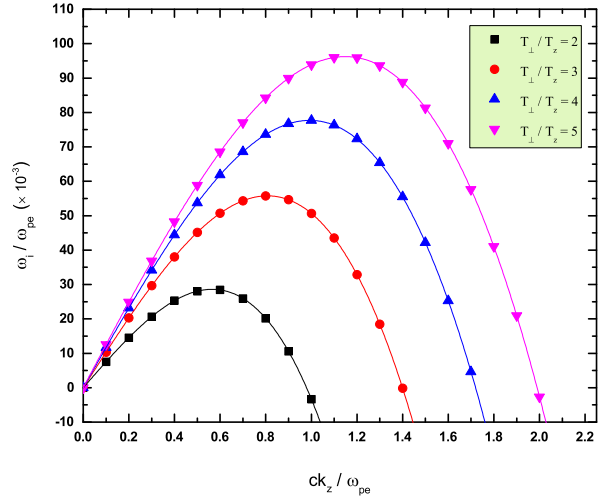}}\caption {The
normalized growth rate of the instability,
$\frac{\omega_{i}}{\omega_{pe}}$ as a function of the wave number,
$\frac{ck_{z}}{\omega_{pe}}$ for different values of the temperature
anisotropy fraction in the fixed $T_{z}=20000 eV$, $n_{0}=10^{32}
m^{-3}$, $\frac{T_{sp}}{T_{z}}=2$ and the magnetic field equal to $8 T$.}
\end{figure}
\clearpage
\begin{figure}
\centerline{ \epsfxsize=10cm \epsffile{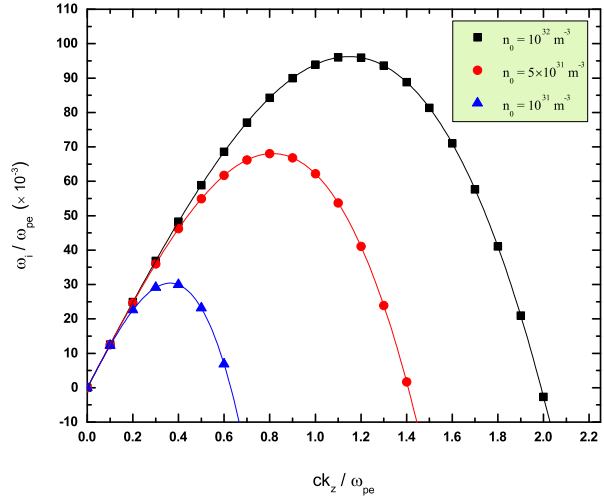}} \caption {The
normalized growth rate of the instability,
$\frac{\omega_{i}}{\omega_{pe}}$ as a function of the wave number,
$\frac{ck_{z}}{\omega_{pe}}$ for different values of the electron
density in the fixed $T_{z}=20000 eV$, $\frac{T\bot}{T_{z}}=5$,
$\frac{T_{sp}}{T_{z}}=2$ and the magnetic field, $10^{8} T$, for the
astrophysical subjects.}
\end{figure}
\clearpage
\end{document}